\documentclass[12pt, draftclsnofoot, journal, a4paper, onecolumn]{IEEEtran}
\usepackage[dvips]{graphicx}
\usepackage{times}
\usepackage{color}
\usepackage{mathrsfs}
\usepackage{amsmath}
\usepackage{amssymb}
\usepackage{cite}
\usepackage{subfigure}

\begin{document}

\newcommand{\mytab}{\hspace*{0.4cm}}
\newcommand{\mytabL}{\hspace*{0.7cm}}
\title{Resource Allocation for Delay Differentiated Traffic in Multiuser OFDM Systems}

\author{Meixia Tao, Ying-Chang Liang and Fan Zhang
\thanks{
This work was presented in part at the IEEE International Conference
on Communications, Istanbul, Turkey, June 2006. The corresponding
author is Meixia Tao, who is with the department of Electrical and
Computer Engineering, National University of Singapore, Singapore
117576. Email: mxtao@nus.edu.sg.}}

\maketitle

\vspace{-1cm}

\begin{abstract}

%
Most existing work on adaptive allocation of subcarriers and power
in multiuser orthogonal frequency division multiplexing (OFDM)
systems has focused on homogeneous traffic consisting solely of
either delay-constrained data (guaranteed service) or
non-delay-constrained data (best-effort service). In this paper, we
investigate the resource allocation problem in a heterogeneous
multiuser OFDM system with both delay-constrained (DC) and
non-delay-constrained (NDC) traffic. The objective is to maximize
the sum-rate of all the users with NDC traffic while maintaining
guaranteed rates for the users with DC traffic under a total
transmit power constraint.
Through our analysis we show that the optimal power allocation over
subcarriers follows a multi-level water-filling principle; moreover,
the valid candidates competing for each subcarrier include only one
NDC user but all DC users.
By converting this combinatorial problem with exponential complexity
into a convex problem or showing that it can be solved in the dual
domain, efficient iterative algorithms are proposed to find the
optimal solutions.
%
%
%
To further reduce the computational cost, a low-complexity
suboptimal algorithm is also developed. Numerical studies are
conducted to evaluate the performance the proposed algorithms in
terms of service outage probability, achievable transmission rate
pairs for DC and NDC traffic, and multiuser diversity.

\end{abstract}

\begin{keywords}
Orthogonal frequency division multiplexing (OFDM), constant-rate
transmission, variable-rate transmission, power control, convex
optimization, water filling.

\end{keywords}

\maketitle

\section{Introduction}
\label{Introduction}

Future broadband wireless networks are expected to support a wide
variety of communication services with diverse quality-of-service
(QoS) requirements. Applications such as voice transmission and
real-time video streaming are very delay-sensitive and need
guaranteed throughput. On the other hand, applications like file
transfer and email services are relatively delay tolerant so
variable-rate transmission is acceptable.
From the physical layer point of view, transmission of
delay-tolerant or non-delay-constrained (NDC) traffic can be viewed
as an ergodic capacity problem \cite{Goldsmith_IT97}, where
maximizing the long-term average transmission rate is the goal.
Thus, wireless resources, such as transmission power and frequency
bandwidth, can be dynamically allocated so as to exploit the time or
frequency selectivities of broadband wireless fading channels.
Likewise, transmission of delay-sensitive or delay-constrained (DC)
traffic can be regarded as a delay-limited capacity problem
\cite{Caire_IT99} in which a constant transmission rate should be
maintained with probability one regardless of channel variations. 
%
In this case, it is desirable to allocate more transmission power
and frequency bandwidth when the channel experiences deep fade and
to allocate less resources when the channel is under favorable
conditions.
%
%
%
%
%
%
We investigate in this work resource allocation in a broadband
wireless network that supports simultaneous transmission of users
with delay differentiated traffic. Our focus is on the formulation
of an analytical framework from the physical layer perspective as
well as the design of efficient and practical algorithms.

%

Multicarrier transmission in the shape of orthogonal frequency
division multiplexing (OFDM) is a leading technique to provide
spectrally efficient modulation as well as user multiplexing in
future wireless systems. With OFDM technique, the broadband wireless
channel is divided into a set of orthogonal narrowband subcarriers.
In a single user system, since the channel frequency responses are
different at different subcarriers, the system performance can be
significantly enhanced by adapting the transmission parameters such
as modulation, coding, and power over each subcarrier. For instance,
the transmitter can send at higher transmission rates over
subcarriers with better channel condition while lower rates or no
data over subcarriers in deep fade. This follows the well-known
water-filling principle.
In a multiuser system, different subcarriers can be allocated to
different users to provide a multiple access method, also known as
OFDMA. As the channels on each subcarrier are likely independent for
different users, the subcarriers experiencing deep fade for one user
may not be in deep fade for other users. As a result, each
subcarrier could be in good condition for some users in a multiuser
OFDM system. By adaptively allocating the subcarriers among multiple
users based on instantaneous channel information, multiuser
diversity can be utilized to boost the overall system efficiency.
%

%
Adaptive resource allocation in multiuser OFDM systems has focused
on homogeneous traffic only. In such systems, the traffic consists
solely of either DC data requiring constant-rate
transmission~\cite{CYWong_99}, or variable-rate NDC data which can
be served in a best-effort manner~\cite{JangLee_03, Peter_ist05,
Ming_wcnc04}.
%
%
For systems with pure DC traffic, the problem is to minimize the
total transmit power while satisfying a basic transmission rate for
each user, which
%
is often referred to as \emph{margin adaption} \cite{Cioffi03}. %
In~\cite{CYWong_99}, an iterative algorithm was proposed to allocate
each user a set of subcarriers and then determine the power and rate
for each user on its allocated subcarriers.
%
%
%
%
For systems with pure NDC traffic, the problem is %
often formulated as maximizing the sum-rate of the system subject to
a total transmit power constraint. This formulation is also known as
\emph{rate adaptation} \cite{Cioffi03}. In \cite{JangLee_03}, it was
shown that the total sum-rate of a multiuser OFDM system is
maximized when each subcarrier is allocated to the user with the
best channel gain for that subcarrier. The total transmit power is
then distributed over the subcarriers using the water-filling
algorithm.
This result holds, however, only for single-antenna systems. It is
no longer optimal when multiple antennas are deployed at the base
station due to the spatial multiplexing gain \cite{Peter_ist05,
Ming_wcnc04, Zhang_05}.
%
Other problem formulations for systems with pure NDC traffic take
user fairness into account. For example, \cite{Cioffi_vtc00} studied
the max-min criterion which aims to maximize the transmission rate
of the bottleneck user. In \cite{Andrews_TW05}, it was proposed to
maintain proportional rates among users for each channel
realization. A utility-function based optimization framework to
balance system efficiency and user fairness was also discussed
in\cite{Li_TW05}.

 In this paper, we consider the subcarrier and power allocation problem
in a heterogeneous multiuser OFDM system where DC and NDC traffic is
supported simultaneously.
%
Users in the system are classified into DC users and NDC users based
on their traffic delay requirements.
%
%
%
We assume that the total transmit power from the base station is
fixed.
Our objective is to maximize the sum-rate of all the NDC users while
maintaining the basic transmission rates of all the DC users over
every transmission frame.
%
%
%
%
%
%
%
%
%
%
%
%
%
%
%
A similar problem was studied in \cite{Anas_04}. However, it assumed
static subcarrier allocation so only the transmit power adaptation
was discussed. Our work, instead, considers joint subcarrier and
power adaptation in multiuser OFDM systems, which is one step
forward of the previous work.
This multiuser subcarrier and power allocation problem is a mixed
integer programming problem, the complexity of which increases
exponentially with the number of subcarriers. To make the problem
more tractable, we transform it into a convex programming problem by
introducing time-sharing variables.
We show that, for a given subcarrier assignment, the optimal power
distribution is achieved by multi-level water-filling.
%
In particular, the water level of each DC user depends explicitly on
the channel gains of its assigned subcarriers and its basic rate
requirement, and can differ from one another. On the other hand, the
water levels of all NDC users are the same.
We also show that, for the optimal subcarrier assignment, the set of
valid user candidates competing for each subcarrier consists of all
the users from the DC group and one from the NDC group with the best
channel gain.
%
%
%
Using these properties, we propose an efficient iterative algorithm
to compute the optimal solution numerically.
Alternatively, the original problem is solved in the dual domain by
using dual decomposition. It is shown that the dual updates can be
done efficiently using an ellipsoid algorithm.
In addition, we present a suboptimal algorithm with linear
complexity in the number of subcarriers and the number of users.

%

The rest of this paper is organized as follows. In Section II we
introduce the system model and describe the problem formulation.
In Section III we formulate the resource allocation problem as a
convex optimization problem by using time-sharing technique and
present analytical frameworks of the optimal solution. An iterative
algorithm to search the optimal solution is also presented. In
Section IV, we attempt to solve the problem using dual approach.
A low-complexity suboptimal algorithm is given in Section V. In
Section VI, we present numerical results of our proposed algorithms
in a multiuser OFDM system. Finally conclusion and discussions are
given in Section VII.

\section{System Model and Problem Formulation}
\label{model}
\setlength\arraycolsep{2pt}

\begin{figure}[htbp]
\begin{center}
\includegraphics[scale = 1]{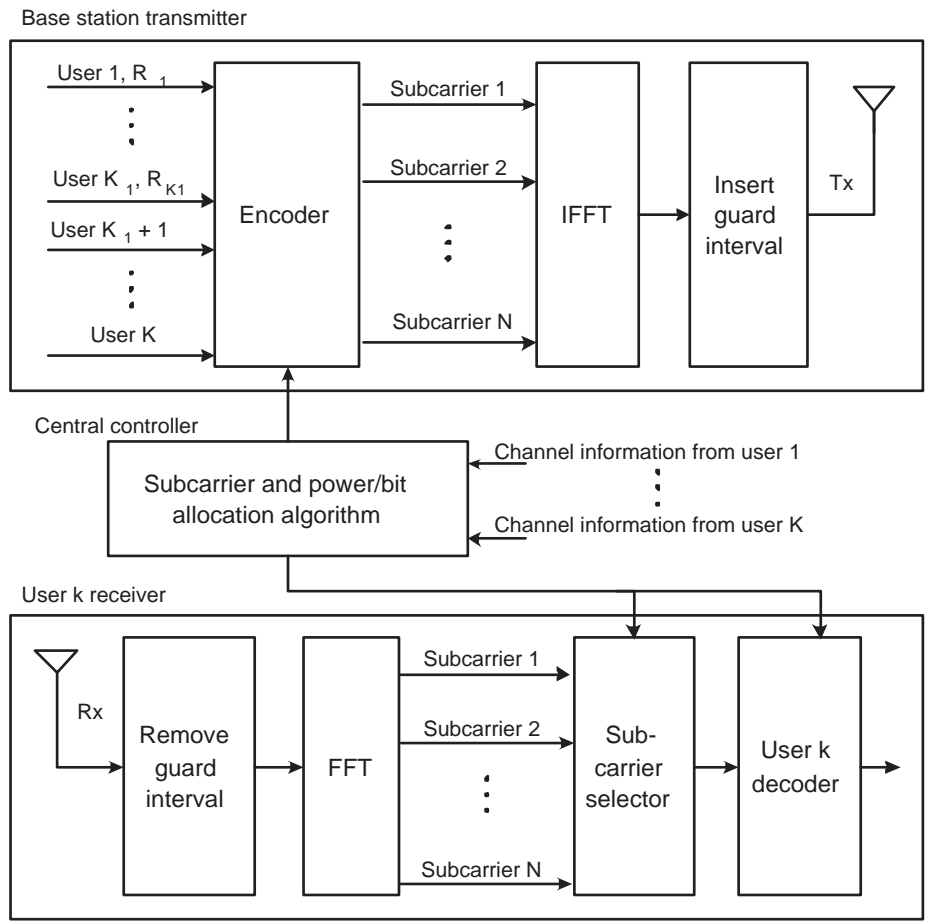}
\caption{Block diagram of a downlink multiuser OFDM system}
\label{fig:model}
\end{center}
\vspace{-0.5cm}
\end{figure}

We consider the downlink of a multiuser OFDM system with block
diagram shown in Fig.~\ref{fig:model}. The system consists of $K$
mobile users. The first $K_1$ users have DC traffic, which requires
a constant transmission rate of $R_k$~($k = 1, \ldots, K_1$) bits
per OFDM symbol, respectively. The traffic of the remaining $K-K_1$
users has no delay constraint and can be delivered in a best-effort
manner.
Note that $K$ is the number of users that are scheduled for
transmission during a certain transmission interval. The total
number of users in a practical system may be much larger than $K$
and, hence, other multiple access techniques such as time-division
multiple access (TDMA) are needed in conjunction with OFDMA.
The data streams from the $K$ users are serially fed into the
encoder block at the base station transmitter.
The total channel bandwidth is $B$ Hz and is divided into $N$
orthogonal subcarriers, which are shared among the $K$ users.
The transmission is on a time-frame basis, where each frame consists
of multiple OFDM symbols.
The fading coefficients of all users are assumed to remain unchanged
within each transmission frame but can vary from one frame to
another.
All channel information is assumed perfectly known at the central
controller, which can be embedded with the base station. Typically,
the channel information can be collected by estimating it at each
user terminal and sending it to the base station via a feedback
channel, or through channel estimation of the uplink in a
time-division duplex system.
Based on the instantaneous channel inputs, the central controller
allocates different subcarriers to different users and determines
the amount of power/bits to be transmitted on each subcarrier
through the subcarrier and power/bit allocation algorithm.
 The resulting allocation information is used to configure the encoder block at the base
station transmitter and to facilitate the subcarrier selector and
decoder at each user receiver. Note that this allocation information
may be sent to each user via a separate channel.
The output data symbols from the encoder are then modulated by the
inverse fast Fourier transform (IFFT). Guard interval is inserted to
ensure orthogonality between OFDM symbols.
%
The total transmit power from the base station is fixed and is given
by $P_T$.

The broadband wireless channel between the base station and each
user terminal is assumed to be frequency-selective Rayleigh fading.
However, the channel in each subcarrier is narrow enough to
experience flat fading.
Let $r_{k,n}$ denote the transmission rate of user $k$ on subcarrier
$n$ in bits per OFDM symbol. It depends on the channel gain
$h_{k,n}$ and the allocated power $P_{k,n}$ of user $k$ on
subcarrier $n$. In general, $r_{k,n}$ can be expressed as
%
%
%
%
\begin{equation}\label{eqn:r_kn}
    r_{k,n} = \log_2{\left(1+\frac{P_{k,n} |h_{k,n}|^2 }{\Gamma
    N_0B/N}\right)},
\end{equation}
where $N_0$ is the power spectral density of additive white Gaussian noise and $\Gamma$ is a
constant, usually called the signal-to-noise ratio (SNR) gap \cite{Cioffi03}.
%
When instantaneous mutual information is used to characterize the
achievable transmission rate, we have $\Gamma = 1$ (0 dB).
If practical signal constellations are used, $\Gamma$ is a constant
related to a given bit-error-rate (BER) requirement. For example,
when uncoded QAM modulation is used we have $\Gamma = -\ln(5\cdot
{\rm BER})/1.5$.
In general, the gap serves as a convenient mechanism for analyzing
the difference between the SNR needed to achieve a certain data rate
for a practical system and the theoretical limit.
Throughout this paper we use (\ref{eqn:r_kn}) as a unified form to
characterize both the theoretical mutual information and practical
transmission rate.

%
%
%
%


The problem we consider here is to optimize the allocation of
subcarriers and power under the total transmit power constraint so
as to maximize the sum-rate of all the $K - K_1$ NDC users while
satisfying the individual rate requirement for each of the $K_1$ DC
users. 
%
Mathematically, the given problem can be formulated as
%
%
\begin{eqnarray} \label{eqn:max_o}
  \max_{\{\Omega_k, P_{k,n}\}} & & \sum_{k=K_1 + 1}^{K}\sum_{n\in
  \Omega_k}{r_{k,n}} \\
%
 {\rm subject~to} && \sum_{n\in \Omega_k}r_{k,n}\ge  R_k, ~k=1,
  \ldots, K_1 \nonumber\\
 && \sum_{k=1}^K \sum_{n\in\Omega_k}{P_{k,n}} = P_T \nonumber\\
 && P_{k,n} \ge 0, \forall k, n \nonumber\\
 && \Omega_1 \cup \Omega_2 \cup \cdots \cup \Omega_K  \subseteq \{1, 2, \ldots,
  N\}\nonumber\\
&& \Omega_1, \Omega_2, \ldots, \Omega_K {~}{\rm are~disjoint}
\nonumber
\end{eqnarray}
%
where $\Omega_k$ is the set of subcarriers assigned to
user $k$.
 $\Omega_k$'s must be mutually
exclusive since each subcarrier is allowed to be used by one user
only.
In general, it is necessary to share the same subcarrier among
multiple users in order to achieve the multiuser capacity
region~\cite{Goldsmith_IT01}. This suggests that superposition
coding together with high complexity decoding should be used.
However, there is only a small range of frequency with overlapping
sharing according to~\cite{Goldsmith_IT01} when optimal power
control is used.
We therefore focus on mutually exclusive subcarrier assignment
schemes, which can also simplify transmitter and receiver
implementation for practical systems.

%
\section{Time-Sharing Based Optimal Subcarrier and Power Allocation}
\label{timeshare}

Finding the optimization variables $\Omega_k$ and $P_{k,n}$ for all
$k$ and $n$ in (\ref{eqn:max_o}) is a mixed integer programming
problem.
In the system with $K$ users and $N$ subcarriers, there are $K^N$
possible subcarrier assignments since each subcarrier can be used by
one user only. For each subcarrier assignment, the total power will
be allocated to meet the individual rate requirement for each DC
user and at the same time to maximize the sum-rate of the NDC users.
The subcarrier assignment together with its associated power
allocation that results in the largest sum-rate while satisfying all
the constraints is the optimal solution.

An approach to make the problem more tractable is to relax the
constraint that each subcarrier is used by one user only. We
introduce a sharing factor $\rho_{k,n} \in [0, 1]$ indicating the
portion of time that subcarrier $n$ is assigned to user $k$ during
each transmission frame. This time-sharing technique was first
proposed in \cite{CYWong_99} and has been frequently used in the
context of subcarrier assignment in multiuser OFDM systems to
convert a mixed integer programming problem into a convex
optimization problem \cite{Cioffi_vtc00, Ming_wcnc04,
Peter_ist05,Andrews_TW05}.
%
In addition, we introduce a variable $s_{k,n}$ and define it as
$s_{k,n}=\rho_{k,n}P_{k,n}$ for all $k$ and $n$. Clearly, $s_{k,n}$
becomes the actual amount of power allocated to user $k$ on
subcarrier $n$, whereas $P_{k,n}$ is the power as if subcarrier $n$
is occupied by user $k$ only. If $\rho_{k,n}=0$, we always have
$s_{k,n}=0$ but $P_{k,n}$ is not necessarily equal to zero.
For notation brevity, we let $\alpha_{k,n} =
|h_{k,n}|^2/(\Gamma_kN_0 B/N)$ for all $k$ and $n$, which is called
the effective \emph{channel-to-noise ratio} (CNR) of user $k$ on
subcarrier $n$. Here, for the purpose of generality, the subindex
$k$ is added to the SNR gap $\Gamma$ to include the case when each
user has different BER requirements if adaptive modulation and
coding is used.
With the aid of time-sharing factors $\rho_{k,n}$ we now readily
transform the original problem (\ref{eqn:max_o}) into:
%
\begin{eqnarray} \label{eqn:max_new}
  \max_{\{\rho_{k,n}, s_{k,n}\}} && \sum_{k=K_1 +
1}^{K}\sum_{n=1}^{N}{ \rho_{k,n}
\log_2{\left(1+\frac{s_{k,n}\alpha_{k,n}}{\rho_{k,n}} \right)}
  }\\
%
%
 {\rm subject~to} && \sum_{n=1}^{N}\rho_{k,n}\log_2{\Big( 1+\frac{s_{k,n} \alpha_{k,n}}{\rho_{k,n}} \Big)}\ge R_k,
   1\le k\le K_1 \label{eqn:R_k}\\
 &&  \sum_{k=1}^K \sum_{n=1}^N{s_{k,n}} = P_T \label{eqn:P_T}\\
 &&  \sum_{k=1}^{K}{\rho_{k,n}} = 1, ~\forall n \label{eqn:rho_1} \\
 &&  s_{k,n} \ge 0, ~ 0\le \rho_{k,n} \le 1, ~\forall k, n~.
 \label{eqn:boundary}
\end{eqnarray}
%
%

%
The objective function (\ref{eqn:max_new}) is a sum of functions of
the form $f(\rho_{k,n}, s_{k,n})=\rho_{k,n}\log_2\big(
1+Cs_{k,n}/\rho_{k,n} \big)$, where $C$ is some positive constant.
By evaluating the Hessian matrix of $f(\rho_{k,n}, s_{k,n})$ at
$\rho_{k,n}$ and $s_{k,n}$, we can prove that $f(\rho_{k,n},
s_{k,n})$ is concave \cite{Boyd_04}. Thus, the objective function is
also concave since any positive linear combination of concave
functions is concave. Moreover, since the inequality constraint
functions in (\ref{eqn:R_k}) are convex and the constraints in
(\ref{eqn:P_T})-(\ref{eqn:boundary}) are all affine, the feasible
set of this optimization problem is convex. Therefore, the problem
in (\ref{eqn:max_new})-(\ref{eqn:boundary}) is a convex optimization
problem and there exists a unique optimal solution, which can be
obtained in polynomial time. In the following we derive some
desirable properties of the optimal solution.

The Lagrangian of the above problem is given by
\begin{eqnarray} \label{eqn:Lag}
  J_1\left(\{\rho_{k,n}\}, \{s_{k,n}\}, \boldsymbol{\beta}, \mu, \boldsymbol{v}\right) &=& \sum_{k=K_1 + 1}^{K}\sum_{n=1}^{N}{\rho_{k,n}\log_2{\left( 1+\frac{s_{k,n}\alpha_{k,n}}{\rho_{k,n}}\right)}}
  + \nonumber\\
  && \sum_{k=1}^{K_1}{\beta_k \left[\sum_{n=1}^N{\rho_{k,n}\log_2{\left( 1 + \frac{s_{k,n}\alpha_{k,n}}{\rho_{k,n}} \right)-R_k} }  \right]} +  \nonumber \\
  && \mu\left(P_T - \sum_{k=1}^K\sum_{n=1}^N{s_{k,n}}\right) +
  \sum_{n=1}^N{v_n \left(1 - \sum_{k=1}^K{\rho_{k,n}} \right)},
\end{eqnarray}
where $\boldsymbol{\beta}=(\beta_1, \ldots, \beta_{K_1}) \succeq 0$,
$\mu \ge 0$ and $\boldsymbol{v} = (v_1, \ldots, v_N)$ are the
Lagrange multipliers for the constraints (\ref{eqn:R_k}),
(\ref{eqn:P_T}) and (\ref{eqn:rho_1}), respectively. The boundary
constraints (\ref{eqn:boundary}) will be absorbed in the
Karush-Kuhn-Tucker (KKT) conditions \cite{Boyd_04} as shown later.
%
%
Let $\rho_{k,n}^*$ and $s_{k,n}^*$ denote the optimal solution, if
it exists, for $1\le k\le K$, $1\le n\le N$. Applying the KKT
conditions, we can obtain the necessary and sufficient conditions
for $\rho_{k,n}^*$ and $s_{k,n}^*$. Specifically, $\rho_{k,n}^*$ and
$s_{k,n}^*$ should satisfy the following equations:
\begin{eqnarray}
  && \frac{\partial J_1(\ldots)}{\partial s_{k,n}^*} \left\{
    \begin{array}{l}
     =0, ~ s_{k,n}^* > 0 \\
     <0, ~ s_{k,n}^* = 0  \\
   \end{array} \right., ~~\forall k,n \label{eqn:partial_P} \\
   && \frac{\partial J_1(\ldots)}{\partial \rho_{k,n}^*} \left\{
    \begin{array}{l}
     <0, ~ \rho_{k,n}^* = 0 \\
     =0, ~ 0< \rho_{k,n}^* < 1  \\
     >0, ~ \rho_{k,n}^* = 1
   \end{array} \right., ~~\forall k,n \label{eqn:partial_rho}
\end{eqnarray}
\begin{eqnarray}
 && \beta_k\left[ \sum_{n=1}^{N}\rho_{k,n}^*\log_2{\Big( 1+\frac{s_{k,n}^* \alpha_{k,n}}
  {\rho_{k,n}^*} \Big)} - R_k \right]= 0, ~~ 1\le k\le K_1. \label{eqn:beta_Rk}
\end{eqnarray}

\subsection{Optimal Power Allocation for Given Subcarrier Assignment}
\label{PA}

In this subsection, we present the optimal power distribution when
subcarrier assignment is given.

Let $\{\rho_{k,n}\}$ be any given subcarrier assignment scheme.
Differentiating the Lagrangian in (\ref{eqn:Lag}) with respect to
$s_{k,n}$ and substituting the result into the KKT condition
(\ref{eqn:partial_P}), we obtain:
%
\begin{eqnarray} \label{eqn:P_opt_dc}
 P_{k,n}^* & = & \frac{s_{k,n}^*}{\rho_{k,n}} = \left( \frac{\beta_k}{\mu \ln 2} - \frac{1}{\alpha_{k,n}} \right)^{+}
\end{eqnarray}
for $k=1, \ldots, K_1$ and $n=1, \ldots, N$, and 
%
%
\begin{eqnarray} \label{eqn:P_opt_ndc}
  P_{k,n}^* &=& \frac{s_{k,n}^*}{\rho_{k,n}}= \left( \frac{1}{\mu \ln2} - \frac{1}{\alpha_{k,n}}
  \right)^{+}
\end{eqnarray}
for $k=K_1+1, \ldots, K$ and $n=1, \ldots, N$.
Here, $(x)^{+} \triangleq \max{(0,x)}$.

Equations (\ref{eqn:P_opt_dc}) and (\ref{eqn:P_opt_ndc}) clearly
show that the optimal power allocation follows the standard
water-filling approach, except that the allocated power is only on
for $\rho_{k,n}$ portion of time. For each user, more power will be
allocated to the subcarriers with higher CNRs and vice versa. But
the water levels of different users can be different. Specifically,
the water level of each DC user is given by $L_k = \beta_k/(\mu \ln
2)$, for $k=1, \ldots, K_1$, and it should ensure the basic rate
requirement $R_k$ in (\ref{eqn:R_k}).
%
%
Substituting (\ref{eqn:P_opt_dc}) into the KKT condition
(\ref{eqn:beta_Rk}) and in view of $\beta_k \ne 0$, we obtain the
closed-form expression for $L_k$ given by:
\begin{eqnarray} \label{eqn:waterlevel_dc}
  L_k &=& \left[ \frac{2^{R_k}}{\prod_{n\in\Omega_k}{(\alpha_{k,n})^{\rho_{k,n}}}}
  \right]^{1/\sum_{n\in\Omega_k}{\rho_{k,n}}},
\end{eqnarray}
where $\Omega_k$ is the set of subcarriers that is assigned to user
$k$ with $\rho_{k,n}>0$ and satisfies $\alpha_{k,n} > 1 / L_k$.
In the case where the given subcarrier assignment is mutually
exclusive, i.e., all the $\rho_{k,n}$'s only take one or zero, the
water levels $L_k$ can be re-expressed as
\begin{eqnarray} \label{eqn:waterlevel_dc_fixsc}
  L_k &=& \left( \frac{2^{R_k}}{\prod_{n'=1}^{g_k}{\alpha_{k,n'}}}
  \right)^{1/g_k},
\end{eqnarray}
where $\alpha_{k,1'}\ge \alpha_{k,2'} \ge \ldots, \ge \alpha_{k,
|\Omega_k|'}$ are the ordered CNRs of the $k$-th DC user on its
allocated subcarrier set $\Omega_k$, and $g_k (\le |\Omega_k|)$ is
the largest integer satisfying $\alpha_{k,g_k'}> 1 / L_k$.
The water level of all the NDC users, on the other hand, is observed
from (\ref{eqn:P_opt_ndc}) to be the same and is given by $L_0 =
1/(\mu \ln 2)$.
%
Let $P_{{\rm DC},T} =
\sum_{k=1}^{K_1}\sum_{n=1}^N{\rho_{k,n}P_{k,n}^*}$ represent the
actual total power consumed by the $K_1$ DC users. Then the water
level of NDC users, $L_0$, can be obtained numerically by using the
total power constraint $P_{{\rm
DC},T}+\sum_{k=K_1+1}^{K}\sum_{n=1}^N{\rho_{k,n}P_{k,n}^*}
  = P_T$. %

\begin{figure}[htbp]
\begin{center}
\includegraphics[scale = 1]{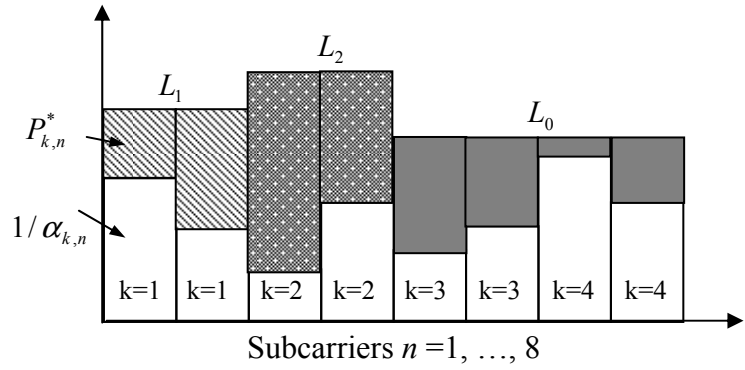}
\caption{Illustration of multi-level water-filling for given
subcarrier assignment in a multiuser OFDM system with 2 DC users, 2
NDC users and 8 subcarriers.} \label{fig:MLWF}
\end{center}
\end{figure}

Fig.~\ref{fig:MLWF} illustrates the optimal power distribution based
on multi-level water-filling in a multiuser OFDM system with $K=4$
users and $N=8$ subcarriers. In the figure, the height of each blank
region represents the inverse of the channel-to-noise ratio and the
height of shadowed ones is the allocated power. The water levels of
the two DC users, $k=1$ and $k=2$, are given by $L_1$ and $L_2$,
respectively, and they are determined explicitly by the basic rate
targets. The water level of the two NDC users, $k=3$ and $k=4$, is
given by $L_0$, and it depends on the total available power after
the subtraction of the power consumed by DC users.
This interpretation on the determination of water levels implicitly
imposes higher priorities on DC users, for which the basic rate
targets must be guaranteed all the time.

\subsection{Optimal Subcarrier Assignment}
\label{OSA} The subcarriers and power should be allocated jointly to
achieve the optimal solution of the problem formulated in
(\ref{eqn:max_new})-(\ref{eqn:boundary}). The previous subsection
discussed the analytical expressions of the optimal power allocation
for a given subcarrier assignment. In this subsection, we derive the
optimal strategy for subcarrier assignment assuming the power
allocation is optimized.

Taking the partial derivative of the Lagrangian in (\ref{eqn:Lag})
with respect to $\rho_{k,n}$, we have:
\begin{eqnarray} \label{eqn:dLdrho}
  \frac{\partial J_1(\ldots)}{\partial \rho_{k,n}} &=&
  {\tilde\beta}_k\left[ \log_2{\Big(1+\frac{s_{k,n}\alpha_{k,n}}{\rho_{k,n}} \Big)}
- \frac{s_{k,n}\alpha_{k,n}}{\ln 2 (\rho_{k,n} +
s_{k,n}\alpha_{k,n})}\right] - v_n,
\end{eqnarray}
where ${\tilde\beta}_k = \beta_k$ for $1\le k \le K_1$ and
${\tilde\beta}_k=1$ otherwise. Now we substitute the optimal power
allocation (\ref{eqn:P_opt_dc}) and (\ref{eqn:P_opt_ndc}) into
(\ref{eqn:dLdrho}) and apply the KKT condition
(\ref{eqn:partial_rho}), then we get:
\begin{eqnarray*}
  \rho_{k,n}^* &=& \left\{ \begin{array}{ll}
    1, & v_n < H_{k,n}(L_0, {\tilde L}_k)  \\
    0, & v_n > H_{k,n}(L_0, {\tilde L}_k)
  \end{array}  \right..
\end{eqnarray*}
Here, the variable ${\tilde L}_k$, for $k=1,\ldots, K$ is defined as
${\tilde L}_k = L_k$ for $1\le k \le K_1$ and ${\tilde L}_k = L_0$
otherwise,
%
%
and the function $H_{k,n}(L_0, {\tilde L}_k)$ is given by:
\begin{eqnarray} \label{eqn:Hkn}
  H_{k,n}(L_0, {\tilde L}_k) &=&
  \frac{{\tilde L}_k}{L_0}\left\{ \left[ \log_2{\big(\alpha_{k,n}{\tilde L}_k \big) } \right]^{+} -
  \frac{1}{\ln 2} \left(1-\frac{1}{\alpha_{k,n}{\tilde L}_k}   \right)^{+}
  \right\}.
\end{eqnarray}
Due to the constraint in (\ref{eqn:rho_1}), we conclude that the
optimal subcarrier assignment is decomposed into $N$ independent
problems. That is,
for each subcarrier $n$, if $H_{k,n}(L_0, {\tilde L}_k)$'s, for
$k=1,\ldots, K$, are all distinct, then only the user with the
largest $H_{k,n}(L_0, {\tilde L}_k)$ can use that subcarrier. In
other words, we have:
\begin{eqnarray} \label{eqn:rho_opt}
  \rho_{k'(n),n}^* = 1, {~} \rho_{k,n}^* = 0, && \forall k\ne k'(n)
\end{eqnarray}
where $k'(n) = \arg\max_{1\le k\le K} {H_{k,n}(L_0, {\tilde L}_k)}$.
%
Hence, it follows that for a given set of water levels $\{L_0,
L_1,\ldots, L_{K_1}\}$, or equivalently, a given set of Lagrange
multipliers $\{\boldsymbol\beta, \mu \}$, we can determine the
optimal subcarrier allocation using (\ref{eqn:rho_opt}). However,
the optimal solution obtained may not satisfy the individual rate
constraint (\ref{eqn:R_k}) and the total power constraint
(\ref{eqn:P_T}).

Function $H_{k,n}(L_0, {\tilde L}_k)$ in (\ref{eqn:Hkn}) plays a key
role in finding the optimal solution of subcarrier and power
allocation. We now take a closer look at its features.
%
%
Firstly, by differentiating $H_{k,n}(L_0, {\tilde L}_k)$ with
respect to $\alpha_{k,n}$ at $\alpha_{k,n}>1/{\tilde L}_k$, it is
seen that $H_{k,n}(L_0, {\tilde L}_k)$ is monotonically increasing
in $\alpha_{k,n}$. As a result, for each subcarrier, the user with a
larger CNR is more likely to be assigned this subcarrier.
In the extreme case where DC users are absent (i.e. $K_1=0$), each
subcarrier will be assigned to the user with the largest CNR. This
agrees with the previous result in \cite{JangLee_03}.
In the general case, the candidate pool for accessing each
subcarrier consists of all the $K_1$ DC users and the only one NDC
user with the largest CNR.
This consequently suggests that the exhaustive search for optimal
subcarrier assignment in the original problem (\ref{eqn:max_o}) has
a complexity of $(K_1 + 1)^N$ rather than $K^N$.
Secondly, we also observe that $H_{k,n}(L_0, {\tilde L}_k)$ is a
non-decreasing function of $L_k$ and $L_0$, respectively, when $1\le
k \le K_1$ and $K_1 < k\le K$. This is obtained by differentiating
$H_{k,n}(L_0, {\tilde L}_k)$ with respect to $L_k$ and $L_0$  for
$1\le k \le K_1$ and $K_1 <k\le K$, respectively. Thus, increasing
the water level $L_k$ of DC user $k$ increases its chance to occupy
more subcarriers. Similarly, increasing the water level $L_0$ of NDC
users allow them to access more subcarriers.
Using the above observations, we propose in the next subsection an
iterative algorithm to find the optimal water levels  and determine
the corresponding subcarrier assignment  so that all the rate and
power constraints are satisfied.
%
%

\subsection{Iterative Algorithm}
The thrust of the algorithm is to obtain the set of optimal water
levels $\{L_0, L_1, \ldots, L_{K_1} \}$ using two nested loops. The
outer loop varies $L_0$ to meet the total power constraint. The
inner loop searches $\{L_1, \ldots, L_{K_1}\}$ and determines the
optimal $\rho_{k,n}$ for all $k$ and $n$ at a given value of $L_0$
to satisfy the basic rate requirement for every DC user.
%
%
%
%
%
%
The algorithm is outlined as follows.

\vspace{0.4cm}
 \hrule
\vspace{0.2cm} \textbf{Optimal Subcarrier Assignment Algorithm}
\vspace{0.2cm} \hrule \vspace{0.3cm}

\emph{\textbf{Main function}}
\begin{itemize}
\item[a)] Set $L_{\rm LB} = L_{\rm UB} = \min_{\substack{K_1<k\le K \\
        1\le n\le N}}{\{1/\alpha_{k,n}\}} - \Delta$\footnote{$\Delta$ is a very small
        number.}
\item[b)] Find the optimal $\{L_k, \rho_{k,n}\}$ at
$L_0=L_{\rm UB}$ \\
~~ Compute the actual power consumption $P_T'$ using (\ref{eqn:P_opt_dc}) and (\ref{eqn:P_opt_ndc}) \\
~~if $P_T' < P_T$, update $L_{\rm UB} = 2 L_{\rm UB}$, and repeat
Step b)
\\
~~else, go to Step c)
\item[c)] Find the optimal $\{L_k, \rho_{k,n}\}$ at $L_0 = (L_{\rm LB}+L_{\rm UB})/2$\\
~~ Compute $P_T'$ using (\ref{eqn:P_opt_dc}) and (\ref{eqn:P_opt_ndc}) \\
~~ if $P_T' > P_T$, let $L_{\rm UB} = L_0$ \\
~~ elseif $P_T' < P_T$, let $L_{\rm LB} = L_0$ \\
~~ Repeat Step c) until $P_T' = P_T$
\end{itemize}

\textbf{\emph{Function}}: find the optimal $\{L_k, \rho_{k,n}\}$ at
a given $L_0$
\begin{itemize}
\item[1)] Set $L_k$ using (\ref{eqn:waterlevel_dc_fixsc}) with $\Omega_k = \{1, \ldots, N\}$, for $k=1,\ldots,
K_1$ \\
~~ Compute $H_{k,n}$ using (\ref{eqn:Hkn}), $\forall k, n$ \\
~~ Obtain $\rho_{k,n}$ using (\ref{eqn:rho_opt}), $\forall k, n$
\item[2)] Compute $R_k' = \sum_{n=1}^N{\rho_{k,n}[\log_2(L_k\alpha_{k,n}) ]^+
}$ for $k=1, \ldots, K_1$
\item[3a)] Find $k^*$ with $R_{k^*}' < R_{k^*}$ and $R_{k^*}' -
        R_{k^*}\le R_{k}' - R_k$ for all $1\le k \le K_1$
\item[3b)] Find the subcarrier set ${\cal A}_{k^*} = \{n | \rho_{k^*,n}
< 1
\}$ for the found $k^*$ \\
~~ for each $n \in {\cal A}_{k^*}$ \\
~~ {\mytab} Let $k(n)={\mathrm{arg}}{\max_{1\le k\le K}}{H_{k,n}}$
and obtain $L_{k^*}(n)$ such that $H_{k^*,n} = (1+\Delta)H_{k(n),n}$
~~
\item[3c)] while $R_{k^*}' < R_{k^*}$ for the found $k^*$ \\
        {\mytab} Denote $m= {\mathrm{arg}}\min_{n\in {\cal A}_{k^*}}{L_{k^*}(n)}$ \\
        {\mytab} Update $L_{k^*}=L_{k^*}(m)$ \\
        {\mytab} Update $\rho_{k^*,m}=1$, $\rho_{k,m}=0$ for $\forall k\ne
        k^*$, and ${\cal A}_{k^*}={\cal A}_{k^*} - \{m\}$ \\
        {\mytab} Compute $R_{k^*}' = \sum_{n=1}^N{\rho_{k^*,n}[\log_2(L_{k^*}\alpha_{k^*,n})
        ]^+}$
 \item[3d)] if $R_{k^*}' > R_{k^*}$, let $R_{k^*}' = R_{k^*}' - \log_2(L_{k^*}\alpha_{k^*,m})$
  \\
 {\mytab} if $R_{k^*}' < R_{k^*}$ \\ 
 {\mytab}{\mytab} Update $\rho_{k^*,m} =
(R_{k^*}-R_{k^*}')/\log_2(L_{k^*}\alpha_{k^*,m})$ and
$\rho_{k(m),m}=1-\rho_{k^*,m}$ \\
{\mytab} elseif $R_{k^*}' > R_{k^*}$ \\
{\mytab}{\mytab} Update $\rho_{k^*,m}=0$, $\rho_{k(m),m}=1$ and
 compute $L_{k^*}$ using (\ref{eqn:waterlevel_dc})
\item[4)] Repeat Steps 2)-3) until $R_{k}'=R_k$ for all $k=1,\ldots, K_1$
\end{itemize}
\vspace{0.2cm} \hrule \vspace{0.4cm}

 In the outer loop (main
function), we initialize $L_0$ to a value slightly below the minimum
reciprocal of the CNRs of all NDC users over all subcarriers so that
none of the NDC users is assigned any power resource based on
(\ref{eqn:P_opt_ndc}). By doing this, the subcarrier and power will
be initially allocated to all DC users as if NDC users were absent.
We then keep increasing $L_0$ until the actual total power
consumption $P_T'$ exceeds the total available power $P_T$ and an
upper bound of $L_0$ is obtained.
Note that if the number of subcarriers in a practical system is
large enough, we can always find at least one subcarrier fully
occupied by one NDC user and, therefore, an analytical upper bound
of $L_0$ can be derived from (\ref{eqn:P_opt_ndc}) as
$L_{\mathrm{UB}} = P_T + \max_{\substack{K_1<k\le K \\ 1\le n\le
N}}\{1/\alpha_{k,n}\}$.
%
%
%
The algorithm then proceeds to use the bisection method to update
$L_0$ so that $P_T'$ converges to $P_T$.
The outer loop converges because the actual total power consumption
$P_T'$ increases monotonically with $L_0$ given the target data
rates for all DC users are satisfied.

%

%
The operation in the inner loop (function: find the optimal $\{L_k,
\rho_{k,n}\}$ at a given $L_0$) is similar to the algorithm
introduced in \cite{CYWong_99}. Each $L_k$ is first initialized to
the minimum water level needed by DC user $k$ to achieve its target
data rate, which happens when all the $N$ subcarriers in the system
are assigned to this user.
%
%
%
%
We then gradually increase $L_k$ for one of the DC users until the
target data rate for this user is satisfied. Increasing $L_k$ is
carried out step by step and the increment at each step is the
minimum required value allowing only one more subcarrier to be added
to this DC user. During this process, there are chances that the
newly added subcarrier is not fully needed and should be time-shared
with other users. It is also possible that this subcarrier is not
needed at all and it only needs to adjust $L_k$ using
(\ref{eqn:waterlevel_dc}).
The algorithm then switches to another DC user and repeats the
process until the target data rates for all DC users are satisfied.
The inner loop converges because for a given DC user $k$, as $L_k$
increases, $H_{k,n}$ for all $n$ increases and more $\rho_{k,n}$
become one. Hence the data rate $R_k' = \sum_{n=1}^N{\rho_{k,n}
[\log_2{(L_k \alpha_{k,n})}]^+}$ increases. On the other hand, the
rates of some other DC users may drop due to their $\rho_{k,n}$
changing from one to zero. Nevertheless, as all the $L_k$'s
increase, the rate of each DC user increases and hence the optimal
$L_k$'s can be approached iteratively.

The complexity of the above algorithm lies in the number of
iterations needed to update $L_0$ in the outer loop and the number
of iterations to update $\{L_k, \rho_{k,n}\}$ for each $L_0$ in the
inner loop. Since $L_0$ is searched by the bisection method, it
requires $\log_2{(1/\epsilon)}$ iterations to converge, where
$\epsilon$ is the accuracy. 
The empirical study in the Appendix shows that the exact number of
iterations to find the optimal $\{L_k, \rho_{k,n}\}$ for each $L_0$
can vary for different values of $L_0$ and different CNR
realizations, but the averaged total number of iterations required
to update the set of water levels $\{L_0, L_1, \ldots, L_{K_1}\}$ in
the whole algorithm can be well approximated by ${\cal
O}(K_1^2/\sqrt{N} \log_2{(1/\epsilon)})$. Since the computational
load in each iteration is linear in $KN$, the overall complexity of
the proposed algorithm is ${\cal O}(K_1^2
KN^{1/2}\log_2{(1/\epsilon)})$.

%

%
%
%

%

%

\subsection{Feasibility and Service Outage}

Similar to the delay-limited capacity problem in information theory
\cite{Caire_IT99}, the constant-rate transmission of DC users
considered in this paper can only be guaranteed in a probabilistic
manner since the total transmit power is fixed and finite. The
service is said to be in an outage if any of the basic rate
requirements cannot be satisfied. Thus, the feasibility of the
optimization problem in (\ref{eqn:max_new})-(\ref{eqn:boundary}) is
directly related to the condition that $P_T \ge P_{{\rm DC}, \min}$,
where $P_{{\rm DC}, \min}$ is the minimum total power needed to
support all $R_k$'s in the absence of NDC users. Finding $P_{{\rm
DC}, \min}$ reduces to the margin adaption problem \cite{CYWong_99}.
%
%
The algorithm proposed in the previous subsection is able to detect
the service outage and obtain the outage probability numerically in
an efficient way. Specifically, if $P_T'$ computed in Step b) of the
main function is greater than $P_T$ when $L_{\mathrm{UB}}$ is still
given by the initial value set in Step a), the algorithm will
terminate and declare an outage.

When outage occurs, one may ignore all the $K_1$ DC users and
allocate all subcarrier and power resources to the $K-K_1$ NDC users
only. Alternatively, one may ignore one or more DC users from the
user list so that at the current channel condition the resources are
sufficient to provide the basic rates of the remaining DC users.
Those ignored DC users may be re-scheduled for transmission at the
next transmission frame by a higher layer scheduler. Further
analysis on higher layer scheduling is out of the scope of this
work.

\section{Subcarrier and power Allocation using Dual Decomposition}

The convex relaxation technique in Section \ref{timeshare} permits
time sharing of each subcarrier. The system model it employs thus
differs from the original OFDMA system where only mutually exclusive
subcarrier assignment is allowed. As a result, the solution gives an
upper bound on the achievable maximum sum-rate of all NDC users
under the individual rate requirement for each DC user and the total
transmit power constraint. Recently, it is shown in \cite{Yu_TCOM06}
that the duality gap of non-convex resource optimization problems in
multicarrier systems is nearly zero if the number of subcarriers is
sufficiently large. Thus, the original problem can be solved in the
dual domain using decomposition method. Applying this result, the
authors in \cite{Cioffi_isit06} developed efficient algorithms to
solve the weighted sum-rate maximization and weighted sum-power
minimization problems in the downlink of multiuser OFDM systems. In
this section we shall apply the result from \cite{Yu_TCOM06} and
solve our original problem (\ref{eqn:max_o}) using the dual
decomposition method. Note that the subcarrier and power allocation
solution in this section provides a lower bound on the maximum
sum-rate of all NDC users when all the target rates for DC users and
the total power constraint are satisfied.

Define $\cal D$ as a set of all non-negative $P_{k,n}$'s for $1\le
k\le K$ and $1\le n\le N$ such that for each $n$ only one $P_{k,n}$
is positive. The Lagrange dual function of the problem
(\ref{eqn:max_o}) is given by:
\begin{eqnarray} \label{eqn:g}
g\left(\boldsymbol{\beta}, \mu\right) &=&\max_{\{P_{k,n} \}\in {\cal D}}J_2\left(\{P_{k,n}\},\boldsymbol{\beta}, \mu \right)\\
 &=& \max_{\{P_{k,n} \}\in {\cal D}}\Bigg[
\sum_{k=K_1+1}^K\sum_{n=1}^N{r_{k,n}} +
\sum_{k=1}^{K_1}{\beta_k}\Big( \sum_{n=1}^Nr_{k,n} - R_k \Big) \nonumber\\
   && +\mu\left(  P_T - \sum_{k=1}^K\sum_{n=1}^NP_{k,n}
   \right)\Bigg], \nonumber
\end{eqnarray}
where $J_2(\{P_{k,n}\}, \boldsymbol\beta, \mu)$ is the Lagrangian
and the dual variables $\{\boldsymbol\beta, \mu \}$ are defined in
the same way as in (\ref{eqn:Lag}).
The dual optimization problem is then formulated as:
\begin{eqnarray}\label{eqn:dual}
  \mathrm{minimize} && { g\left( \boldsymbol{\beta},\mu \right)}\\
 \mathrm{subject~to} && \boldsymbol{\beta} \succeq 0, \mu \ge 0. \nonumber
\end{eqnarray}
Note that the Lagrangian $J_2(\ldots)$ is linear in $\beta_k$ and
$\mu$ for fixed $P_{k,n}$, and  $g\left(\boldsymbol{\beta},
\mu\right)$ is the maximum of these linear functions, so the dual
problem (\ref{eqn:dual}) is convex.

To solve the dual problem, we first decompose the dual function into
$N$ independent optimization problems:
\begin{eqnarray} \label{eqn:g_dec}
 g\left(\boldsymbol{\beta}, \mu\right) &=& \sum_{n=1}^N{g_n\left(\boldsymbol{\beta}, \mu\right)} -
\sum_{k=1}^{K_1}{\beta_k R_k} + \mu P_T,
\end{eqnarray}
where
\begin{equation}\label{eqn:gn}
    g_n\left(\boldsymbol{\beta}, \mu\right) = \max_{\{P_{k,n} \}\in {\cal D}}
    \left[ \sum_{k=K_1+1}^Kr_{k,n} + \sum_{k=1}^{K_1}\beta_k r_{k,n} - \mu \sum_{k=1}^KP_{k,n}
    \right].
\end{equation}
Suppose subcarrier $n$ is assigned to user $k$. Using the KKT
condition similar to (\ref{eqn:partial_P}), the optimal $P_{k,n}^*$
that maximizes the object of the max operation in (\ref{eqn:gn}) for
fixed $\boldsymbol{\beta}$ and $\mu$ can be readily obtained and is
given by (\ref{eqn:P_opt_dc}) when $1\le k \le K_1$, or
(\ref{eqn:P_opt_ndc}) when $K_1 < k \le K$. 
%
Substituting (\ref{eqn:P_opt_dc}) and (\ref{eqn:P_opt_ndc}) into
(\ref{eqn:gn}) and comparing all the $K$ possible user assignments
of this subcarrier, we obtain
\begin{equation}\label{eqn:gn_max}
    g_n\left(\boldsymbol{\beta}, \mu\right) = \max_{1\le k\le K}
  \left\{\tilde{\beta}_k \left[ \log_2{\left(\frac{\alpha_{k,n}\tilde{\beta}_k }{\mu \ln2}  \right)}\right]^{+} -
  \mu \left( \frac{\tilde{\beta}_k}{\mu \ln2}   -\frac{1}{\alpha_{k,n}}   \right)^{+}
  \right\},
\end{equation}
where $\tilde{\beta}_k = \beta_k$ for $k=1, \ldots, K_1$ and
$\tilde{\beta}_k=1$ for $k=K_1+1,\ldots, K$.

%
%
%

Once (\ref{eqn:gn_max}) is solved for all $n$'s, the dual function
$g\left(\boldsymbol{\beta}, \mu\right)$ can be obtained using
(\ref{eqn:g_dec}). Since it is convex, a gradient-type algorithm can
 minimize $g\left(\boldsymbol{\beta}, \mu\right)$ by
updating $\{\boldsymbol{\beta}, \mu\}$ simultaneously along some
appropriate search directions, which is guaranteed to converge to
the optimal solution. In general, $g\left(\boldsymbol{\beta},
\mu\right)$ is not differentiable, and thus its gradient does not
exist. Nevertheless, we can resort to the subgradient derived in the
following proposition.

\emph{Proposition 1}: For the dual problem (\ref{eqn:dual}) with
primal defined in (\ref{eqn:max_o}), the following is a subgradient
of $g\left(\boldsymbol{\beta}, \mu\right)$
\begin{eqnarray*}
  \Delta\beta_k &=& \sum_{n=1}^N{r_{k,n}^*} - R_k, ~ k=1\ldots, K_1, \\
  \Delta\mu &=& P_T - \sum_{k=1}^N\sum_{n=1}^NP_{k,n}^*,
\end{eqnarray*}
where $P_{k,n}^*$ maximizes the Lagrangian $J_2(\ldots)$ over $\cal
D$ at $\boldsymbol\beta$ and $\mu$, and
$r_{k,n}^*=\log_2{(1+P_{k,n}^* \alpha_{k,n})}$.
%
%

\begin{proof} By definition of $g\left(\boldsymbol{\beta},
\mu\right)$ in (\ref{eqn:g}):
\begin{eqnarray*}
  g\left(\boldsymbol{\beta}',\mu'\right) & \ge &
  \sum_{k=K_1+1}^{K}\sum_{n=1}^N{r_{k,n}^*} + \sum_{k=1}^{K_1}{
\beta_k' \left( \sum_{n=1}^N{r_{k,n}^*} -R_k\right)}
+ \mu'\left( P_T - \sum_{k=1}^K\sum_{n=1}^N {P_{k,n}^*} \right)  \\
   &=& g\left(\boldsymbol{\beta},\mu\right) + \sum_{k=1}^{K_1}{
(\beta_k' - \beta_k) \left( \sum_{n=1}^N{r_{k,n}^*} -R_k\right)}
+(\mu'-\mu) \left( P_T - \sum_{k=1}^K\sum_{n=1}^N {P_{k,n}^*}
\right).
\end{eqnarray*}
Proposition 1 is hence proven by using the definition of
subgradient.
\end{proof}
With the above subgradient, both the subgradient and ellipsoid
methods \cite{Boyd_note06} can be used to update
$\{{\boldsymbol\beta}, \mu \}$. Here we choose the ellipsoid method
 which converges in ${\cal{O}}((K_1+1)^2)$ iterations. The algorithm details can be found in \cite{Boyd_note06}.
The following lemma leads to a suitable choice of the initial
ellipsoid.

\emph{Lemma 1}:  The optimal dual variables $\{
\boldsymbol{\beta}^*, \mu^*\}$ must satisfy
\begin{eqnarray*}
  0 &\le& \mu^* \le \mu^{\max} = \frac{1}{\ln{2}} \max_{\substack{K_1<k\le K \\ 1\le n\le N}
  }{\{\alpha_{k,n}\}},
\label{eqn:mu_ub} \\
  0 &\le& \beta_k^* \le  \beta_k^{\max} = \max_{\substack{K_1<k\le K \\ 1\le n\le N}
  }{\{\alpha_{k,n}\}} \left[P_T + \frac{1}{\min_{1\le n\le N}{\{\alpha_{k,n}\}}}
  \right]. \label{eqn:beta_ub}
\end{eqnarray*}
\begin{proof}
The dual variables $\{\boldsymbol{\beta}^*,\mu^*\}$ must satisfy the
KKT conditions in order to be optimal. Taking the partial derivative
of $J_2(\ldots)$ in (\ref{eqn:g}) with respect to $P_{k,n}$ results
in
\begin{eqnarray}\label{eqn:P2proof1}
  \frac{\alpha_{k,n}}{\ln{2} \left(1 + \alpha_{k,n}P_{k,n} \right)} &=&
  \mu,
\end{eqnarray}
if user $k$, for $K_1 < k \le K$, is active in subcarrier $n$, or
\begin{eqnarray}\label{eqn:P2proof2}
  \frac{\beta_k\alpha_{k,n}}{\ln{2} \left(1 + \alpha_{k,n}P_{k,n} \right)} &=&
  \mu,
\end{eqnarray}
if user $k$, for $1 \le k \le K_1$, is active in subcarrier $n$.
Since $P_{k,n}$ must always satisfy $0\le P_{k,n}\le P_T$ due to the
power constraint, we obtain the upper bound $\mu^{\max}$ by letting
$P_{k,n} = 0$ in (\ref{eqn:P2proof1}) and the upper bound
$\beta_k^{\max}$ by substituting $\mu^{\max}$ into
(\ref{eqn:P2proof2}) and letting $P_{k,n} = P_T$.
\end{proof}

Using Lemma 1, one may choose an initial ellipsoid that encloses the
hyper-cuboid where $\{\boldsymbol{\beta}^*, \mu^* \}$ resides,
namely, $E(\mathbf{A}_0, \mathbf{z}_0) = \{ \mathbf{x} |
(\mathbf{x}-\mathbf{z}_0)^T \mathbf{A}_0^{-1}
(\mathbf{x}-\mathbf{z}_0) \le 1\}$,
where
\begin{eqnarray*}
\mathbf{A}_0 & = & \mathrm{diag}\left[
\begin{array}{cccc}
                      (1+K_1)\Big(\frac{1}{2}\beta_1^{\max}\Big)^2, & \cdots, & (1+K_1)\Big(\frac{1}{2}\beta_{K_1}^{\max}\Big)^2, &
                      (1+K_1)\Big(\frac{1}{2}\mu^{\max}\Big)^2
                    \end{array}
 \right] \\
  \mathbf{z}_0 &=& \left[\begin{array}{cccc}
                      \frac{1}{2}\beta_1^{\max}, & \cdots, & \frac{1}{2}\beta_{K_1}^{\max}, &
                      \frac{1}{2}\mu^{\max}
                    \end{array} \right]^T.
\end{eqnarray*}

Due to duality gap, after obtaining the optimal dual variables
$\{\boldsymbol{\beta}^*, \mu^* \}$ that minimize the dual function,
it remains to find the optimal primal solutions $\{P_{k,n}^*\}$ that
maximize the Lagrangian $J_2(\ldots)$ and satisfy all the rate and
power constraints in the original problem (\ref{eqn:max_o}).
We can solve this by first identifying the subcarrier assignment
$\{\Omega_k^*\}$ using (\ref{eqn:gn_max}) with
$\{\boldsymbol{\beta}^*, \mu^* \}$ substituted and then determining
the power allocation $\{P_{k,n}^* \}$ using the results derived in
Section III-A.

%

\section{Suboptimal Subcarrier and Power Allocation} \label{subopt}
%

%

%
%
%
%
%


In this section we present a suboptimal allocation algorithm that
 has a much lower computational cost compared with
both the optimal iterative algorithm in Section III-C and the dual
update algorithm in Section IV.
The idea is to first obtain the subcarrier assignment for the DC
users by assuming that the power is equally distributed over all
subcarriers and that all the NDC users are absent. After that, the
power distribution for each DC user over its assigned subcarrier set
is individually refined. The purpose of the refinement is to
minimize the power while maintaining the basic transmission rates.
At last, the residual subcarriers and power are distributed among
the NDC users using the optimal resource allocation algorithm in
\cite{JangLee_03} to maximize the sum-rate.
This algorithm is suboptimal because the subcarrier assignment for
DC users in the first step is obtained by assuming equal power
allocation. The decoupling of subcarrier assignment and power
allocation for DC users carried out in the first two steps, though
being suboptimal, can greatly simplify the complexity and is often
used for resource allocation in multiuser OFDM systems such as
\cite{Cioffi_vtc00, Andrews_TW05, Kim_vtc04}
%
%
%

The outline of the proposed suboptimal subcarrier assignment scheme
for DC users is presented below.
%
%

%

\vspace{0.4cm} \hrule \vspace{0.2cm} \textbf{Suboptimal Subcarrier
Allocation Algorithm for DC users} \vspace{0.2cm} \hrule
\vspace{0.2cm}

\begin{enumerate}
    \item
        set $R_k' = 0$, $\Omega_k = \emptyset$ for all $k=1,\ldots,
        K_1$ and ${\cal A} = \{1, 2, \ldots, N \}$
    \item
%
    while ${\cal A} \ne \emptyset$ and $R_k' < R_k$ for any $1\le k \le K_1$
\begin{itemize}
       \item[a)]find $k^*$ with $R_{k^*}' < R_{k^*}$ and $R_{k^*}' -
        R_{k^*} \le R_{k}' - R_k$ for all $1\le k \le K_1$
        \item[b)]for the found $k^*$, find $n^*$ satisfying $\alpha_{k^*,n^*} \ge \alpha_{k^*,n}$
        for $n\in {\cal A}$
        \item[c)]for the found $k^*$ and $n^*$, update $\Omega_{k^*} = \Omega_{k^*} \cup \{n^*\}$, ${\cal A} = {\cal A} - \{ n^*\}$
            and $R_{k^*}' = R_{k^*}' + \log_2{( 1 + \frac{\alpha_{k^*,n^*}P_T}{N}
            )}$
\end{itemize}
\end{enumerate}
\vspace{0.2cm} \hrule \vspace{0.4cm}

%
%
At each iteration of Step $2)$ in the above algorithm, the DC user
whose current data rate $R_k'$ is the farthest away from its target
rate $R_k$ will be allowed to pick a new subcarrier from the
available subcarrier set. Preferably, the subcarrier with the
highest CNR will be chosen.
Upon acquiring $\Omega_k$ for $1\le k\le K_1$, the power
distribution for each DC user is then adjusted using the analytical
solution (\ref{eqn:P_opt_dc}) and (\ref{eqn:waterlevel_dc_fixsc}).
%
%
%
In the case where $g_k < |\Omega_k|$ for some $k$, the above
suboptimal subcarrier assignment algorithm over-allocates
subcarriers to DC user $k$. To efficiently utilize all the wireless
resources, the remaining $|\Omega_k| - g_k$ subcarriers should be
returned to the residual subcarrier set $\cal A$, which will be
distributed among the $K-K_1$ NDC users.
%
%
%
%
%
 Let $P_{{\rm DC}, T}$ denote the actual
power consumption of all the $K_1$ DC users. If $P_{{\rm DC}, T}$ is
larger than the total power limit $P_T$, a service outage occurs.
Otherwise, the residual transmit power $P_T- P_{{\rm DC},T}$
together with the residual subcarrier set $\cal A$ are allocated
over the $K - K_1$ NDC users. Specifically, each subcarrier in
${\cal A}$ is assigned to the NDC user with the highest CNR, and the
power is distributed over these subcarriers in the form of
water-filling (\ref{eqn:P_opt_ndc}), where the water level can be
determined by $P_T - P_{{\rm DC},T}$.

The number of iterations involved in finding the suboptimal
$\Omega_k$'s for $k=1, \ldots, K$ is limited by $N$ since $N$ is the
total number of subcarriers available. That is, the proposed
suboptimal algorithm only performs a fixed number of iterations
rather than iterating till convergence.
The power allocation for given $\{\Omega_k\}$ has explicit
analytical solution as shown in Section III-A and its complexity is
linear in $KN$. Therefore, the overall complexity of this suboptimal
algorithm is only linear in $K$ and $N$.
%
%

\section{Numerical Results}
In this section we present numerical performance results of the
proposed optimal and suboptimal resource allocation algorithms.
We consider a multiuser OFDM system with $N=64$ subcarriers and
$K=8$ users. Therein, $K_1 = 4$ users have DC traffic and the others
 have NDC traffic. For simplicity, we let the rate
requirements of all DC users be identical and equal to $R_{\rm
DC}/K_1$ bits/OFDM symbol, where $R_{\rm DC}$ denotes the sum of the
basic rates.
In all simulations, the channel from the base station to each user
terminal is modeled by the HiperLan/2 channel model A
\cite{ETSI_98}, which is an 8-tap channel with exponential power
delay profile, 20MHz sampling frequency and 50ns rms
(root-mean-square) delay spread. The channels for different users
are assumed to be independent.
%
We also assume that the path losses from the base station to all
 user terminals are the same.
The average channel gain on each subcarrier is normalized.
%
The system total transmit SNR is defined as $P_T/(N_0B)$.
The SNR gap in the rate function (\ref{eqn:r_kn}) is set to 6.6 (8.2
dB) for both DC users and NDC users. In practice, when uncoded QAM
constellation is used the SNR gap of 8.2 dB corresponds to a BER
requirement of $10^{-5}$.

%

%
To evaluate the performance of the three proposed adaptive resource
allocation algorithms, we also present the results for two
non-adaptive schemes in comparison. In both schemes, the subcarrier
assignment is predetermined but the power allocation for each user
over its predetermined set of subcarriers follows the optimal
approach derived in Section \ref{PA}. In the first scheme, all the 8
users are treated equally and each is assigned 8 subcarriers. We
refer to this scheme as Fixed Subcarrier Assignment with Optimal
Power Allocation (FSA-OPA).
In the other scheme, DC users are given higher priority than NDC
users and each is assigned 12 subcarriers, whereas each NDC user is
allocated 4 subcarriers only. This scheme is called Fixed Subcarrier
Assignment with \emph{Priority} and with Optimal Power Allocation
(FSAP-OPA).
%
%
%
%
%
%
In addition, for both schemes, we let the predetermined subcarriers
for each user spread over the entire bandwidth in a comb pattern
\cite{MUOFDM_99}. This prevents the situation where all subcarriers
of a user are in deep fade.
%

%
%
%
%
%

\begin{figure}[t!]
\begin{center}
\includegraphics[scale = 0.76]{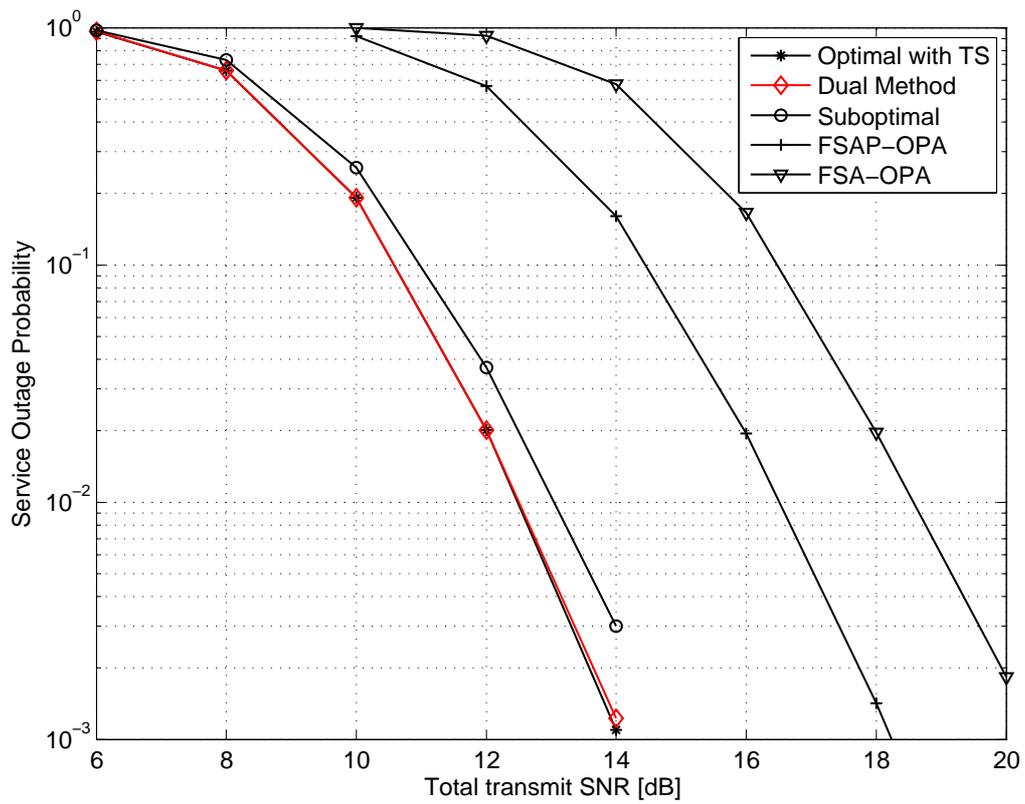}
\caption{Service outage probability versus total transmit SNR at
$R_{\rm DC}=80$ bits/OFDM symbol.} \label{fig:outage_snr}
\end{center}
\end{figure}

%
We first compare the performance in terms of service outage
behavior.
Fig.~\ref{fig:outage_snr} illustrates the service outage probability
versus total transmit SNR when the total target transmission rate of
DC users is $R_{\rm DC}=80$ bits/OFDM symbol.
It is first observed that the time-sharing based optimal algorithm
and the dual method perform almost identically. This suggests that
two algorithms result in almost the same subcarrier assignment
solution for DC users. This observation is expected because the
duality gap vanishes when $N$ is sufficiently large and, as a
result, both the upper bound given by the optimal algorithm with
time sharing and the lower bound given by the dual method approach
the truly optimal solution.
One can also see that the performance loss due to the suboptimal
subcarrier assignment in the suboptimal algorithm is marginal. In
particular, at a service outage probability of $1\%$, the SNR loss
is within $0.5$ dB.
In addition, it is seen from Fig.~\ref{fig:outage_snr} that the
proposed adaptive algorithms significantly outperform
 the two fixed subcarrier assignment (FSA) schemes. At
moderate and high SNR regions, the service outage probability is
more than an order of magnitude lower.
Besides, the FSA scheme with priority outperforms the one without
priority as more subcarriers are assigned to DC users in the former.

\begin{figure}[t!]
\begin{center}
\includegraphics[scale = 0.76]{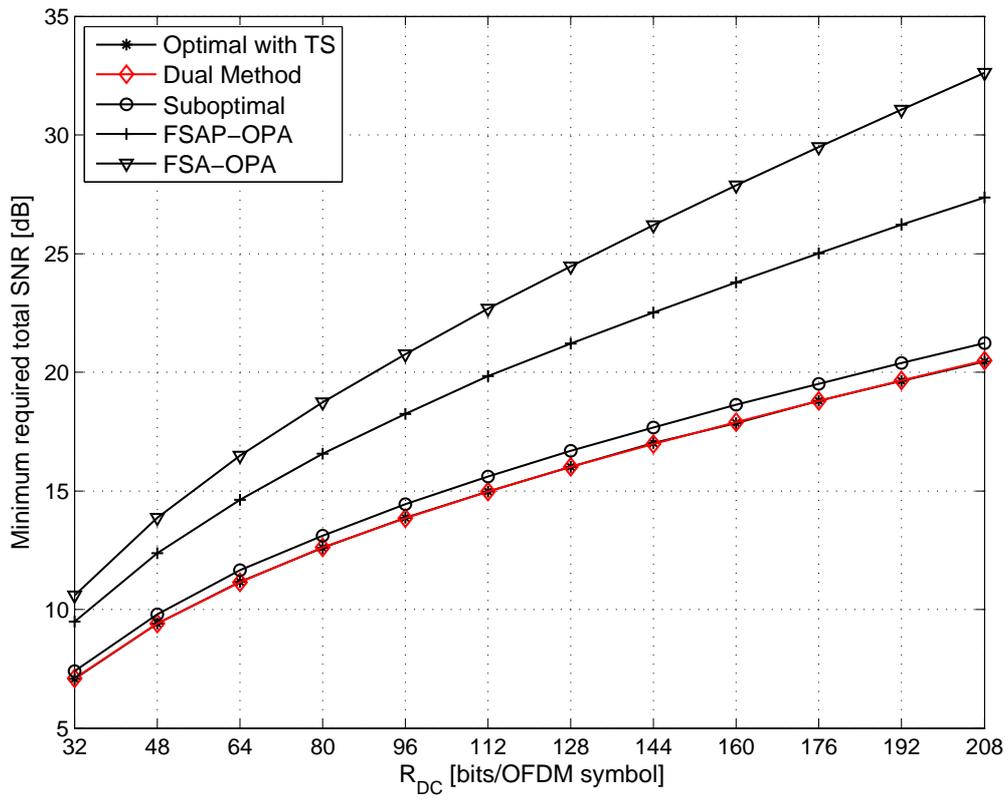}
\caption{Minimum required total transmit SNR versus $R_{\rm DC}$ at
a service outage probability of $1\%$.} \label{fig:outage}
\end{center}
\end{figure}

In Fig.~\ref{fig:outage}, we plot the minimum required total
transmit SNR for different $R_{\rm DC}$ at a given service outage
probability of $1\%$.
It is again observed that the optimal algorithm with time sharing
and the dual method have almost identical performance. Therefore,
only the results of the former will be illustrated hereafter.
From Fig.~\ref{fig:outage} we observe that, for a wide range of
$R_{\rm DC}$ that the multiuser OFDM system can support with $1\%$
outage probability,
the difference on the minimum required SNR between the optimal and
suboptimal algorithms is consistently less than $0.5$ dB. In
particular, as $R_{\rm DC}$ decreases, the performance of the
suboptimal algorithm becomes closer to that of the optimal
algorithm. This is expected as the suboptimality of the proposed
suboptimal algorithm lies only in the subcarrier allocation for DC
users. If the rate requirement for DC users is small, the suboptimal
algorithm will become nearly optimal.
%
Fig.~\ref{fig:outage} also shows that, as $R_{\rm DC}$ increases,
the minimum required total SNR of the proposed adaptive algorithms
increases at a much lower speed than that of the two FSA schemes.
%
%
%

\begin{figure}[htbp]
\begin{center}
\includegraphics[scale = 0.76]{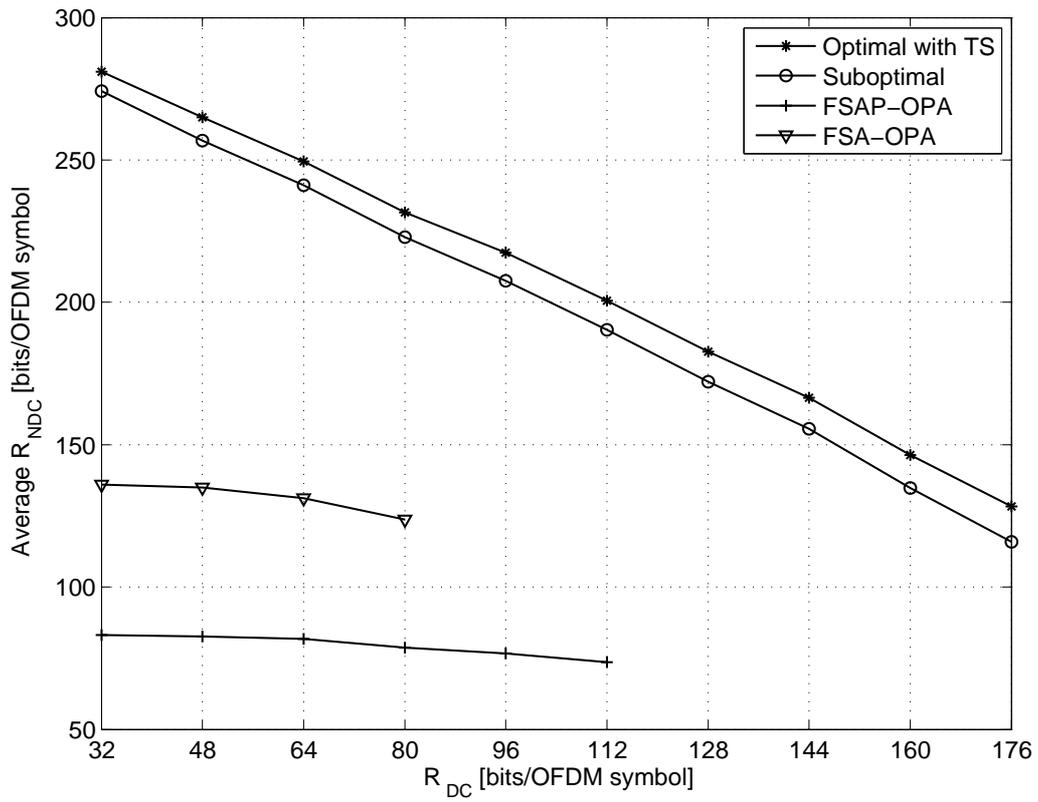}
\caption{Achievable ($R_{\rm DC}$, ${\bar R}_{\rm NDC}$) rate pairs
at a total transmit SNR of 20 dB.} \label{fig:rate_region}
\end{center}
\end{figure}

%
We next study the achievable transmission rates of the heterogenous
multiuser OFDM system with the proposed adaptive subcarrier and
power allocation algorithms.
Fig.~\ref{fig:rate_region} shows the achievable pairs of the basic
sum-rate for DC traffic $R_{\rm DC}$ and the average sum-rate for
NDC traffic ${\bar R}_{\rm NDC}$
%
%
at a total transmit SNR of $20$ dB. The average sum-rate for NDC
traffic ${\bar R}_{\rm NDC}$ is obtained by averaging the
instantaneous sum-rates of NDC users over $500$ independent channel
realizations.
To ensure a service outage probability of $1\%$ or below, the
maximum value of $R_{\rm DC}$ in our simulation is set to $176$
bits/OFDM symbol for the proposed algorithms, and to $112$ and $80$
bits/OFDM symbols, respectively, for FSAP-OPA and FSA-OPA.
%
%
%
The maximum achievable $R_{\rm DC}$ with an acceptable service
outage probability, for example $1\%$, at a given total SNR can be
obtained from Fig.~\ref{fig:outage}.
From Fig.~\ref{fig:rate_region} one can observe that, compared with
the optimal subcarrier and power allocation algorithm, the loss of
the average achievable NDC traffic rate at a given $R_{\rm DC}$ by
using the suboptimal algorithm is within $2\%\sim 9\%$. On the other
hand, compared with the two FSA schemes, both the proposed optimal
and suboptimal adaptive algorithms demonstrate substantially larger
achievable rate regions.
%
%
%
We also observe that, at the same $R_{\rm DC}$, the ${\bar R}_{\rm
NDC}$ of FSA-OPA is larger than that of FSAP-OPA. This is because
NDC users have fewer subcarriers in FSAP-OPA. However, the maximum
$R_{\rm DC}$ FSAP-OPA can support is larger than that of FSA-OPA.

\begin{figure}[htbp]
\begin{center}
\includegraphics[scale = 0.76]{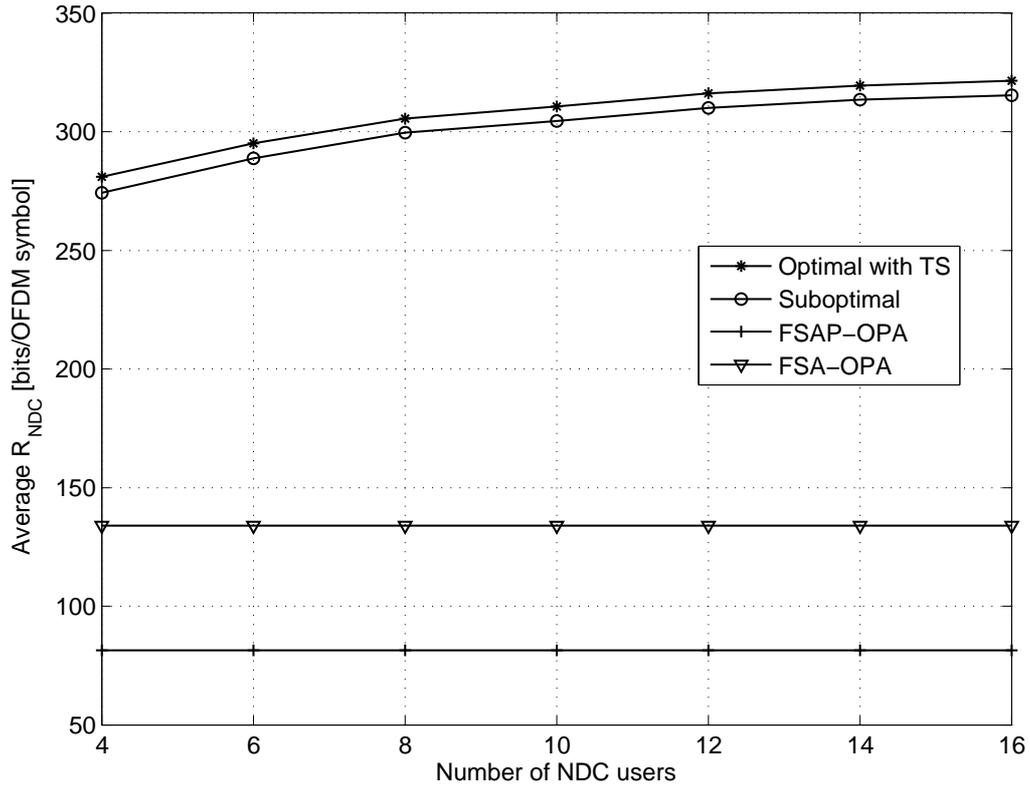}
\caption{Achievable ${\bar R}_{\rm NDC}$ versus the number of NDC
users at a total transmit SNR of 20 dB with $4$ DC users.}
\label{fig:MUD}
\end{center}
\end{figure}

Finally, we demonstrate the multiuser diversity exploited by our
algorithms. We let the number of DC users in the system be fixed at
$K_1=4$ and vary the number of NDC users between $4$ and $16$.
Fig.~\ref{fig:MUD} presents the average sum-rate ${\bar R}_{\rm
NDC}$ as a function of the number of NDC users at $R_{\rm DC}=32$
bits/OFDM symbols. Same to Fig.~\ref{fig:rate_region}, the total
transmit SNR is $20$ dB and ${\bar R}_{\rm NDC}$ is obtained by
averaging $500$ independent channel realizations. In the two FSA
schemes, the subcarrier allocation for DC users is the same as
before, but the rest of the subcarriers are all allocated to one NDC
user, which is selected in a round-robin fashion at each
transmission frame. The values of ${\bar R}_{\rm NDC}$ for the two
FSA schemes remain constant since no multiuser diversity is
achieved. On the contrary, ${\bar R}_{\rm NDC}$ obtained by the
proposed adaptive algorithms increases as the number of NDC users
increases, which clearly shows the multiuser diversity. In
particular, the achievable ${\bar R}_{\rm NDC}$ of the optimal
algorithm is about $110\%$ and $140\%$ higher than that of the
FSAP-OPA scheme when the system has $4$ and $16$ NDC users,
respectively.

%

%
%

%
%
%
%

%

\section{Conclusion and Discussions}

Supporting communication services with diverse QoS requirements in
future broadband wireless networks is crucial. This paper considered
the resource allocation problem in an OFDM based downlink system
that supports simultaneous transmission of users with DC traffic at
constant rates and users with NDC traffic at variable rates. We
investigated this problem from the physical layer perspective and
aimed to maximize the sum-rate of NDC traffic while maintaining
individual basic rates of DC traffic for each channel realization
under a total power constraint.
%
%
%
%
%
%
%
%
It was shown that the optimal power allocation over the subcarriers
in such a heterogeneous system has the form of multi-level
water-filling;
%
moreover, the set of valid user candidates competing for each
subcarrier consists of only one NDC user but all DC users.
We converted this combinatorial problem with exponential complexity
into a convex problem using the time-sharing technique and developed
an efficient iterative algorithm with polynomial complexity. We also
solved the original problem using dual decomposition method which
leads to polynomial complexity as well.
To further speed up the resource allocation and make it more
suitable for practical systems, we then proposed a suboptimal
algorithm whose computation load is only linear in the number of
users and subcarriers in the system.
The performance of our algorithms was evaluated in terms of service
outage probability, achievable DC and NDC traffic rate pairs, and
multiuser diversity. The numerical results showed that the convex
relaxation technique with time sharing and the dual decomposition
approach obtained almost the same solution and that the suboptimal
algorithm has the near optimal performance.
Results also demonstrated that the proposed adaptive subcarrier and
power allocation algorithms significantly outperform the schemes
with adaptive power allocation but fixed subcarrier assignment.

This paper adopted the continuous rate function (\ref{eqn:r_kn}),
which greatly helped to derive the insights of optimal resource
allocation. If discrete rates are used in practical systems, our
algorithms can be modified accordingly. In particular, since the
proposed suboptimal algorithm has near-optimal performance at
significantly lower complexity, it is more desirable to modify the
suboptimal one. For instance, one can obtain the subcarrier
assignment using the proposed suboptimal algorithm and then apply
the greedy bit loading algorithm for each single user as in
\cite{CYWong_99}. Nevertheless, our continuous rate formulation
provides the performance upper bound for systems with discrete
rates.

We have also assumed that the channels from the base station to all
the NDC users have the same path loss. By symmetry, our formulation
also equalizes the long-term average throughput among all the NDC
user. To achieve fairness when their channel path losses are
different, we can simply modify our cost function (\ref{eqn:max_o})
by dividing the channel-to-noise ratio with the path loss. By doing
so, the effective channel gains for all NDC users are normalized.
Therefore, only the user whose current channel condition is at its
peak level will be selected to compete with DC users for each
subcarrier. This is similar to the concept of ``riding on the
channel peak" in opportunistic scheduling.
%
%

\appendix[Empirical Study on the convergence speed of the iterative
algorithm in Section III-C]

\begin{figure}[htbp]
\begin{center}
\includegraphics[scale = 0.65]{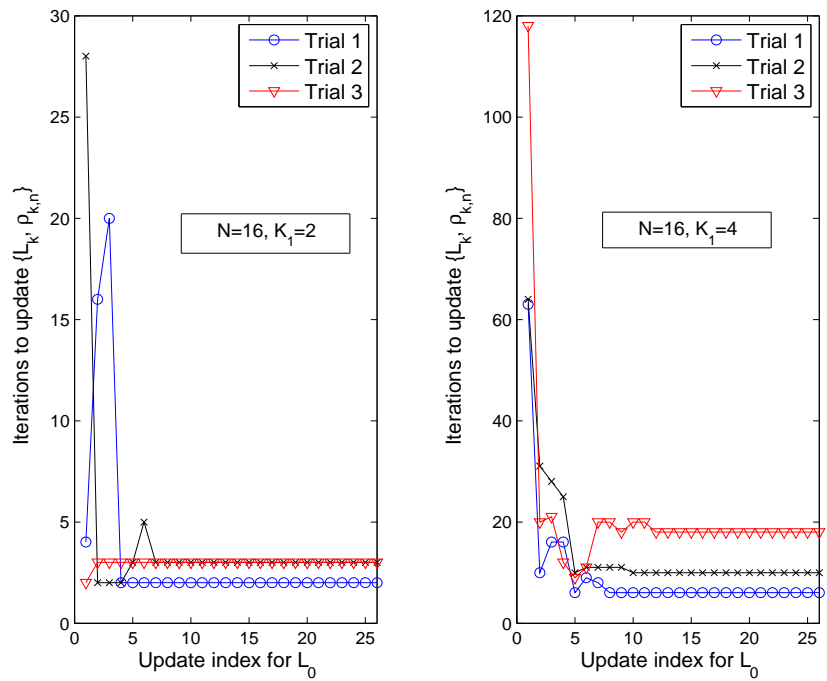}
\caption{Iterations required to update $\{L_k, \rho_{k,n} \}$ at
each $L_0$ update. } \label{fig:snapshot_ite}
\end{center}
\end{figure}

\begin{figure}[htbp]
\begin{center}
\includegraphics[scale = 0.76]{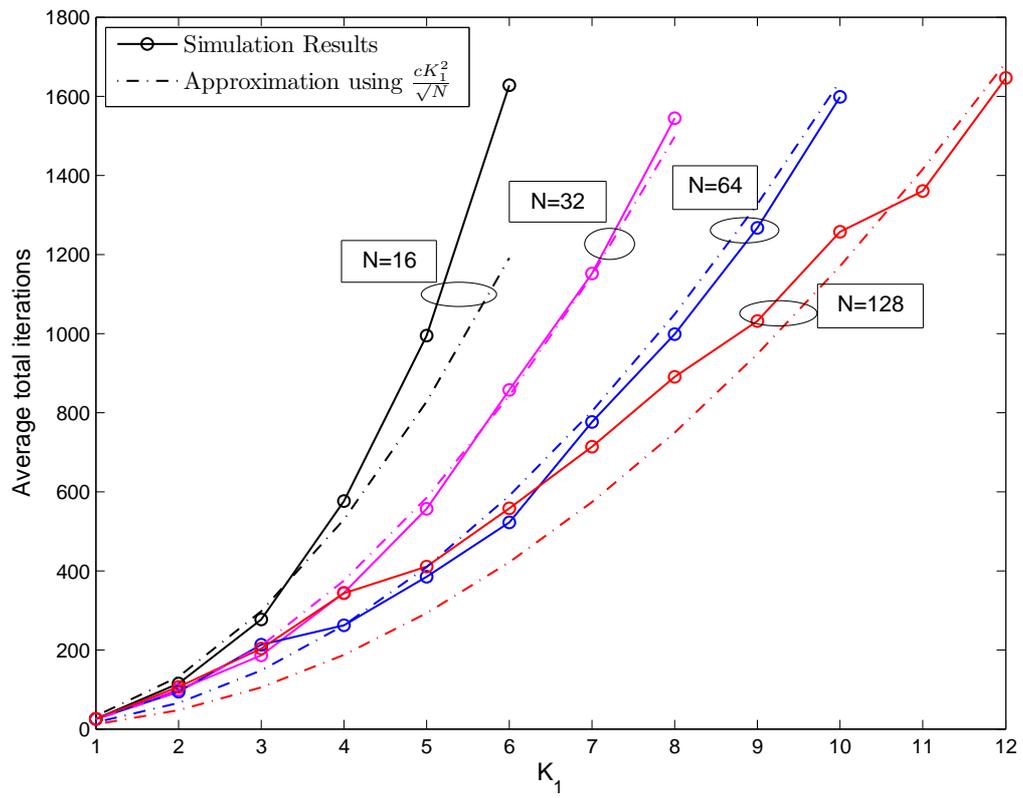}
\caption{ Total iterations required to update $\{L_0, L_k,
\rho_{k,n} \}$.} \label{fig:avg_ite}
\end{center}
\end{figure}

Simulation settings: $P_T = 100$, $2$ NDC users, $R_k = 16$ bits/
OFDM symbol for $1\le k \le K_1$, $K_1\in \{1, \ldots, 12\}$,
$N\in\{16, 32, 64, 128\}$. $\alpha_{k,n} = N |h_{k,n}|^2$ are
randomly generated with $h_{k,n}$ modeled as complex Gaussian
variables of zero mean and unit variance and independent for all $k$
and $n$. The accuracy of bisection-searching $L_0$ is set to
$\epsilon=10^{-7}$, and it leads to $26$ iterations in the main
function throughout this simulation study. The parameter $\Delta$ in
Step 3b) of the inner function is set to $0.005$. Note that it
typically takes very few iterations in the while loop of Step 3c) in
the inner function to find the $L_{k^*}$ for DC user $k^*$ that
meets its rate requirement. Thus we choose to count the number of
times it repeats for Steps 2)-3) as the number of iterations to
update $\{L_k, \rho_{k,n}\}$ at a given $L_0$.
Fig.~\ref{fig:snapshot_ite} shows the snapshot of iterations to
update $\{L_k, \rho_{k,n}\}$ at each updating step of $L_0$ for
three random channel realizations. The number of iterations varies
for different $L_0$ and different channel realizations and, in
general, more iterations are needed when $K_1$ increases.
To extract the rules on how the number of iterations change with $N$
and $K_1$, we plot in Fig.~\ref{fig:avg_ite} the averaged total
iterations needed to find the set of optimal solutions $\{L_0, L_k,
\rho_{k,n}\}$, where each value is obtained by averaging over $20$
independent channel realizations. For comparison we also plot the
curves generated using the analytical expression $c K_1^2/\sqrt{N}$
with the constant $c$ being $c=26\times 5.1=132.6$. It is observed
that the analytical expression provides a very good approximation on
the shape of the simulated curves. Therefore, we conclude that the
proposed time-sharing based optimal subcarrier assignment algorithm
converges in ${\cal O} (K_1^2/\sqrt{N} \log_2{(1/\epsilon)})$
iterations.



\bibliographystyle{IEEEtran}
\bibliography{reference}

\begin{thebibliography}{10}
\providecommand{\url}[1]{#1}
\csname url@rmstyle\endcsname
\providecommand{\newblock}{\relax}
\providecommand{\bibinfo}[2]{#2}
\providecommand\BIBentrySTDinterwordspacing{\spaceskip=0pt\relax}
\providecommand\BIBentryALTinterwordstretchfactor{4}
\providecommand\BIBentryALTinterwordspacing{\spaceskip=\fontdimen2\font plus
\BIBentryALTinterwordstretchfactor\fontdimen3\font minus
  \fontdimen4\font\relax}
\providecommand\BIBforeignlanguage[2]{{%
\expandafter\ifx\csname l@#1\endcsname\relax
\typeout{** WARNING: IEEEtran.bst: No hyphenation pattern has been}%
\typeout{** loaded for the language `#1'. Using the pattern for}%
\typeout{** the default language instead.}%
\else
\language=\csname l@#1\endcsname
\fi
#2}}

\bibitem{Goldsmith_IT97}
A.~J. Goldsmith and P.~Varaiya, ``Capacity of fading channels with channel side
  information,'' \emph{IEEE Trans. on Infor. Theory}, vol.~43, no.~6, pp.
  1986--1992, Nov. 1997.

\bibitem{Caire_IT99}
G.~Caire, G.~Taricco, and E.~Biglieri, ``Optimal power control over fading
  channels,'' \emph{IEEE Trans. on Infor. Theory}, vol.~45, no.~5, pp.
  1468--1489, July 1999.

\bibitem{CYWong_99}
C.~Y. Wong, R.~S. Cheng, K.~B. Letaief, and R.~D. Murch, ``Multiuser {OFDM}
  with adaptive subcarrier, bit and power allocation,'' \emph{IEEE Journal on
  Selected Areas in Comm.}, vol.~17, no.~10, pp. 1747--1758, Oct. 1999.

\bibitem{JangLee_03}
J.~Jang and K.~B. Lee, ``Transmit power adaptation for multiuser {OFDM}
  systems,'' \emph{IEEE Journal on Selected Areas in Comm.}, vol.~21, no.~2,
  pp. 171--178, Feb. 2003.

\bibitem{Peter_ist05}
P.~W.~C. Chan and R.~S. Cheng, ``Optimal power allocation in zero-forcing
  {MIMO-OFDM} downlink with multiuser diversity,'' in \emph{Proc. of IST Mobile
  \& Wireless Communications Summit}, Dresden, June 2005.

\bibitem{Ming_wcnc04}
Y.~M. Tsang and R.~S. Cheng, ``Optimal resouce allocation in {SDMA/MIMO/OFDM}
  systems under {QoS} and power constraints,'' in \emph{Proc. of IEEE WCNC},
  2004.

\bibitem{Cioffi03}
J.~M. Cioffi, ``Digital communications,'' \emph{EE379 Course Reader, Stanford
  University}, 2003.

\bibitem{Zhang_05}
Y.~J. Zhang and K.~B. Letaief, ``An efficient resource allocation scheme for
  spatial multiuser access in {MIMO/OFDM} systems,'' \emph{IEEE Trans. Comm.},
  vol.~53, no.~1, pp. 107--116, Jan. 2005.

\bibitem{Cioffi_vtc00}
W.~Rhee and J.~M. Cioffi, ``Increase in capacity of multiuser {OFDM} system
  using dynamic subchannel allocation,'' in \emph{Proc. of IEEE VTC}, 2000.

\bibitem{Andrews_TW05}
Z.~Shen, J.~G. Andrews, and B.~L. Evans, ``Adaptive resource allocation in
  multiuser {OFDM} systems with proportional fairness,'' \emph{IEEE Trans. on
  Wireless Comm.}, vol.~4, no.~6, pp. 2726--2737, Nov. 2005.

\bibitem{Li_TW05}
G.~Song and Y.~Li, ``Cross-layer optimization for {OFDM} wireless networks:
  {Part II}: Algorithm development,'' \emph{IEEE Trans. on Wireless Comm.},
  vol.~4, no.~2, pp. 625--634, March 2005.

\bibitem{Anas_04}
M.~Anas, K.~Kim, S.~Shin, and K.~Kim, ``{QoS} aware power allocation for
  combined guaranteed performance and best effort users in {OFDMA} systems,''
  in \emph{Proc. International Symposium on Intelligent Signal Processing and
  Comm. Systems}, Nov. 2004.

\bibitem{Goldsmith_IT01}
A.~J. Goldsmith and M.~Effros, ``The capacity region of broadcast channels with
  intersymbol interference and colored {Gaussian} noise,'' \emph{IEEE Trans. on
  Infor. Theory}, vol.~47, pp. 219--240, 2001.

\bibitem{Boyd_04}
S.~Boyd and L.~Vandenberghe, \emph{Convex Optimization}.\hskip 1em plus 0.5em
  minus 0.4em\relax Cambridge University Press, 2004.

\bibitem{Yu_TCOM06}
W.~Yu and R.~Lui, ``Dual methods for nonconvex spectrum optimization of
  multicarrier systems,'' \emph{IEEE Trans. on Comm.}, vol.~54, no.~7, pp.
  1310--1322, July 2006.

\bibitem{Cioffi_isit06}
K.~Seong, M.~Mhoseni, and J.~M. Cioffi, ``Optimal resource allocation for
  {OFDMA} downlink systems,'' in \emph{Proc. of ISIT'06}, Seattle, USA, July
  2006.

\bibitem{Boyd_note06}
S.~Boyd, ``Convex optimization ii,'' \emph{EE364B Course Note, Stanford
  University. Available at http://www.stanford.edu/class/ee364b/}, 2006.

\bibitem{Kim_vtc04}
K.~Kim, H.~Kang, and K.~Kim, ``Providing quality of service in adaptive
  resource allocation for {OFDMA} systems,'' in \emph{Proc. of IEEE VTC'04},
  vol.~3, May 2004, pp. 1612--1615.

\bibitem{ETSI_98}
E.~N. Committee, ``Channel models for {HIPERLAN/2} in different indoor
  scenarios,'' \emph{European Telecom. Standards Institute, Sophia-Antipolis,
  Valbonne, France, Norme ETSI, doc 3ERI085B}, 1998.

\bibitem{MUOFDM_99}
E.~Lawrey, ``Multiuser {OFDM},'' in \emph{Proc. of International Symposium on
  Signal Processing and its Applications (ISSPA'99)}, Brisbane, Australia, Aug.
  1999.

\end{thebibliography}

\newpage


\newpage

\end{document}